\documentclass[12pt]{article}
\usepackage{cite}

\usepackage{amsmath,amsfonts,amssymb}
\usepackage{graphicx,color}
\usepackage{epsfig,epsf}
\usepackage{bm}

\usepackage{float}
\usepackage[section]{placeins} 
\usepackage{relsize}

\usepackage[bookmarks, breaklinks, colorlinks,urlcolor=black, citecolor=red, 
linkcolor=blue]{hyperref}

\usepackage{url}
\usepackage{subfigure}
\usepackage{multirow}

\definecolor{Darkgreen}{RGB}{30,150,30}

\newcommand{\nn}{\nonumber}
\def\gev{\ensuremath{\mathrm{\,Ge\kern -0.1em V}}}
\def\sn#1{\textit{Scenario-#1}}

\begin{document}
\vspace{-1cm}

\hspace*{5.3in}{\small{IITH-PH-0004/20}}\\
\hspace*{5.55in}{\small {SI-HEP-2020-12}}
\vspace*{1mm}
\renewcommand*{\thefootnote}{\fnsymbol{footnote}}

\begin{center}
	{\Large\bf{	Feeble neutrino portal dark matter at neutrino detectors}}
	\\[6mm]
	{Priyotosh Bandyopadhyay $^1$\footnote{Email: bpriyo@phy.iith.ac.in}, 
		Eung Jin Chun $^2$\footnote{Email: ejchun@kias.re.kr}, 
		Rusa Mandal $^3$\footnote{Email: Rusa.Mandal@uni-siegen.de}},
	\\[3mm]
	
	$^1${\small\em Indian Institute of Technology Hyderabad, Kandi,  Sangareddy-502287, Telengana, India}\\
	$^2${\small\em Korea Institute for Advanced Study, Seoul 02455, Korea}\\
	$^3${\small\em Theoretische Physik 1, Naturwissenschaftlich-Technische Fakult$\ddot{a}$t, \\ Universit$\ddot{a}$t Siegen, 57068 Siegen, Germany}
\end{center}

\begin{abstract}
	
We explore the neutrino portal dark matter (DM) at its minimum of field content having two dark sector particles coupled to a right-handed neutrino.  Assuming the feeble nature of their interactions with the standard model (SM) particles,  we analyze the freeze-in production of the observed DM relic density characterized by three different categories depending on the major production mechanisms. The portal provides interesting signatures at the neutrino detectors like KamLAND, Super-Kamiokande and IceCube, from a very late decay of the scalar DM to the fermion DM and the SM neutrino. Such neutrino flux spectrum from the cosmic and galactic origins can produce anomalous signals at future experiments.

\end{abstract}
\setcounter{footnote}{0}
\renewcommand*{\thefootnote}{\arabic{footnote}}

\vspace*{-1cm}

\section{Introduction}

The past decades have seen the proliferation of ideas on the identity of dark matter (DM) and the origin of its abundance which requires physics beyond the standard model (SM). The SM needs also to be extended to generate tiny neutrino mass, which is plausibly linked with the presence of DM. For instance, a right-handed neutrino (RHN) could be a mediator to the dark sector \cite{Pospelov:2007mp}. The role of RHN as an arbiter in attaining the observed DM relic density via the standard freeze-out mechanism has been extensively studied  \cite{Falkowski:2009yz, Gonzalez-Macias:2016vxy,Escudero:2016ksa,Tang:2016sib, Campos:2017odj,Batell:2017rol,Blennow:2019fhy,Hall:2019rld}. 
This conveniently assumes a neutrino Yukawa coupling large enough to maintain the RHN in thermal equilibrium. Sometimes it necessitates to have a very small coupling resulting in non-thermalized RHN
and thus delayed freeze-out of the DM  \cite{Bandyopadhyay:2011qm, Bandyopadhyay:2017bgh,Dror:2016rxc,Okawa:2016wrr,Kopp:2016yji,Bandyopadhyay:2018qcv}. Yet another way of thermal generation of the DM abundance can be achieved even with an extremely tiny (Dirac) neutrino Yukawa coupling \cite{Asaka:2005cn} which is generalized to the freeze-in mechanism applicable to various situations \cite{McDonald:2001vt,  Hall:2009bx}.

In this article, we present a general study of freeze-in production in neutrino-portal DM. We assume all the dark sector particles as well as the RHN are feebly interacting with the SM sector and thus all of them are never in thermal equilibrium and their abundances  are produced via various freeze-in processes.  We categorize the freeze-in mechanism into three representative scenarios depending on the major process determining the final DM abundance satisfying the observed relic density. 
For some related studies, we refer the readers to  \cite{Becker:2018rve,Chianese:2018dsz,Bian:2018mkl,Chianese:2019epo,Cosme:2020mck}.  

While all the conventional direct and indirect DM detections are obsolete in the freeze-in scenarios of  neutrino-portal DM, there can exists a dark sector particle  which decays very late to a neutrino and a stable DM component, and thus leaves an interesting signal in the neutrino detectors \cite{PalomaresRuiz:2007ry,Cui:2017ytb} if it decays by now.  We generalize this consideration to our dark sector particle which has a wide range of lifetime, and analyze the energetic neutrino spectrum observable in neutrino flux measurements. 
Another interesting impact arises from the late-time injection of neutrinos contributing to dark radiation (DR)   which is constrained by the CMB measurements \cite{planck18}.  In a recent study of multi-component DM scenario where a fraction of the DM components decays to neutrinos,  imposes bounds on the product of that fraction and the lifetime of a decaying dark matter \cite{Poulin:2016nat}, which  are applicable to our analysis. 
These two considerations provide complimentary limits on the model parameter space.  We remark that direct neutrino detection at KamLAND~\cite{KamLAND}, Super-Kamiokande (SK) \cite{superk,Bays:2011si} and IceCube \cite{icecube} provides fairly strong bounds for some region of the parameter space. 

The article is arranged as follows.  In Sec.~\ref{sec:relic} we present the model and three different scenarios of the freeze-in mechanism. Potential signatures of the model are discussed and  constraints from neutrino flux measurements at the neutrino detectors and CMB data are obtained in Sec.~\ref{sec:DarkRad}. We conclude in Sec.~\ref{sec:concl}. Appendix~\ref{sec:appendix} contains the expressions of all the cross-sections and decay rates involved in the calculation.

\section{Freeze-in via feeble neutrino/Higgs portal } \label{sec:relic}

In the extension of the SM we introduce two dark sector particles, a Majorana fermion $\chi$ and a real scalar $\phi$ which have a portal coupling $\lambda$ with a RHN $N$:
\begin{align}
\label{eq:Lag}
-\mathcal{L}_{\rm new} \subset  \left\{   \lambda N \chi \phi  + y_\nu LHN+ {\rm h.c.} \right\}
+ \kappa \phi^2 |H|^2\,,
\end{align}
where we omitted the mass terms of $N$, $\chi$ and $\phi$ which will be denoted by $m_N$, $m_\chi$ and $m_\phi$, respectively.
Note that the RHN has the neutrino Yukawa coupling $y_\nu$ with the SM lepton and Higgs doublets, respectively denoted by $L$ and $H$, realizing the Type-I seesaw mechanism. 
We implicitly assume that  the observed neutrino masses and mixing are induced by two RHNs other than the portal RHN $N$ whose contribution is negligible due to its feeble coupling. 
 The scalar $\phi$ has also a Higgs portal coupling $\kappa$. 
To ensure the stability of the dark sector, an extra $Z_2$ symmetry has been invoked  under which the dark sector fields are odd.
After the electroweak symmetry breaking, $H=(v+h)/\sqrt{2}$, the Higgs portal coupling contributes to the mass term for the scalar $\phi$, and induces the $h$-$\phi$-$\phi$ coupling $\kappa v$. 
 We assume $m_\chi< m_\phi$ and thus $\chi$ is the primary DM candidate. 
However, in addition, $\phi$ can also be a  (decaying) DM candidate if it lives long enough.

 In the subsequent discussions we explore the interplay of the three couplings,
 $y_\nu, \kappa$ and $\lambda$, in explaining the observed DM abundance through the freeze-in mechanism.  Due to the tiny $y_\nu$ coupling of $N$ and the feebly interacting nature of the dark sector particles $\chi$ and $\phi$, none of these three new particles are assumed ever in equilibrium with the thermal bath of the SM particles throughout our study. Before going into the detailed analysis, it is useful to get an estimate of the couplings ensuring the freeze-in nature. Requiring all the dominant production rate of these particles  to remain smaller than the expansion rate of the Universe, $\Gamma < H(T) \sim T^2/M_{pl}$ at temperature $T$, one finds
\begin{align*}
\Gamma_{N \to \nu h} & \approx \frac{y_\nu^2}{8\pi} m_N  < H(T) \big|_{T=m_N}
 \Longrightarrow y_\nu \lesssim 10^{-7}\,, \\
\Gamma_{\phi\phi \to h h} &\approx \frac{\kappa^4}{4\pi} T < H(T) \big|_{T=m_\phi} 
 \Longrightarrow \kappa \lesssim 10^{-7}\,, \\
\Gamma_{\phi\chi \to \ell h} &\approx \frac{y_\nu^2 \lambda^2}{4\pi} T <H(T) \big|_{T=m_{\phi,\chi}} \Longrightarrow y_\nu\lambda \lesssim 10^{-7}\,, 
\end{align*}
where we used the reduced Planck mass $M_{pl}=2.44 \times 10^{18}$ GeV, and $m_{N,\phi,\chi}\sim 1 $\,TeV.

Assuming such feeble couplings, the freeze-in abundances of $N, \chi$ and $\phi$ will be determined by solving relevant Boltzmann equations with the initial condition $ Y_N=Y_{\phi}=Y_{\chi}=0$.
The coupled Boltzmann equations for the system of $N,\chi,\phi$, and the SM particles in the thermal bath are given by  
\begin{align}
\label{eq:dYDM}
\frac{d Y_\chi}{dx}=+& \frac{1}{x^2} \frac{s(m_\chi)}{H(m_\chi)} \langle \sigma v \rangle_{NN \to \chi\chi} \left(\!Y_N^2 -\! \left(\frac{Y_N^{\text{eq}}}{Y_\chi^{\text{eq}}}\right)^{\!\!\!2}Y_\chi^2\!\right) + \frac{1}{x^2} \frac{s(m_\chi)}{H(m_\chi)} \langle \sigma v \rangle_{\phi\phi\to\chi\chi}\! \left(\!Y_\phi^2 -\! \left(\!\frac{Y_\phi^{\text{eq}}}{Y_\chi^{\text{eq}}}\right)^{\!\!\!2}\!Y_\chi^2\!\right)\! \nn \\
-& \frac{1}{x^2} \frac{s(m_\chi)}{H(m_\chi)} \langle \sigma v \rangle_{\chi\phi \to h \nu}\!\left(Y_\chi Y_\phi - Y_\chi^{\text{eq}} Y_\phi^{\text{eq}} \right) + \frac{\tilde{\Gamma}_{\phi \to \chi N}}{H(m_\chi)}\, 
x \left(\!Y_\phi - \frac{Y_\phi^{\rm eq}}{Y_\chi^{\rm eq} Y_N^{\rm eq}} \, Y_\chi Y_N\!\right) \nn \\
+& \frac{\tilde{\Gamma}_{\phi \to \chi \nu}}{H(m_\chi)} 
\,x \left(Y_\phi - \frac{Y_\phi^{\rm eq}}{Y_\chi^{\rm eq} }  Y_\chi \right) + \frac{\tilde{\Gamma}_{N \to \chi \phi}}{H(m_\chi)} 
\,x \left(Y_N - \frac{Y_N^{\rm eq}}{Y_\chi^{\rm eq} Y_\phi^{\rm eq}}  Y_\chi Y_\phi\right), \\
\label{eq:dYphi}
\frac{d Y_\phi}{dx}= -& \frac{1}{x^2} \frac{s(m_\chi)}{H(m_\chi)} \langle \sigma v \rangle_{\phi\phi\to\chi\chi}\! \left(\!Y_\phi^2 -\! \left(\!\frac{Y_\phi^{\text{eq}}}{Y_\chi^{\text{eq}}}\right)^{\!\!\!2}Y_\chi^2\right) 
- \frac{1}{x^2} \frac{s(m_\chi)}{H(m_\chi)} \langle \sigma v \rangle_{\phi\phi\to NN} \left(\!Y_\phi^2 \!-\! \left(\frac{Y_\phi^{\text{eq}}}{Y_N^{\text{eq}}}\right)^{\!\!\!2} Y_N^2\!\right) \nn \\
-&\frac{1}{x^2} \frac{s(m_\chi)}{H(m_\chi)} \langle \sigma v \rangle_{\phi\phi\to {\rm SM}} \left(Y_\phi^2 - {Y_\phi^{\text{eq}}}^2\right) - \frac{1}{x^2} \frac{s(m_\chi)}{H(m_\chi)} \langle \sigma v \rangle_{\chi\phi \to h \nu}\!\! \left(Y_\chi Y_\phi \!-\!Y_\chi^{\text{eq}} Y_\phi^{\text{eq}} \right)  \nn \\
-& \frac{\tilde{\Gamma}_{\phi \to \chi N}}{H(m_\chi)} 
\,x \left(Y_\phi - \frac{Y_\phi^{\rm eq}}{Y_\chi^{\rm eq} Y_N^{\rm eq}}  Y_\chi Y_N\right) - \frac{\tilde{\Gamma}_{\phi \to \chi \nu}}{H(m_\chi)} 
\,x \left(Y_\phi - \frac{Y_\phi^{\rm eq}}{Y_\chi^{\rm eq} }  Y_\chi \right)  \nn \\
 + & \frac{\tilde{\Gamma}_{N \to \chi \phi}}{H(m_\chi)} 
\, x \left(Y_N - \frac{Y_N^{\rm eq}}{Y_\chi^{\rm eq} Y_\phi^{\rm eq}}  Y_\chi Y_\phi\right)\,, \\
\label{eq:dYN}
\frac{d Y_N}{dx} = -&\frac{1}{x^2} \frac{s(m_\chi)}{H(m_\chi)} \langle \sigma v \rangle_{NN \to \chi\chi}\!\! \left(\!\!Y_N^2 \!-\! \left(\frac{Y_N^{\text{eq}}}{Y_\chi^{\text{eq}}}\right)^{\!\!\!2}Y_\chi^2\!\right)
+ \frac{1}{x^2} \frac{s(m_\chi)}{H(m_\chi)} \langle \sigma v \rangle_{\phi\phi\to NN} \left(\!Y_\phi^2 \!-\! \left(\frac{Y_\phi^{\text{eq}}}{Y_N^{\text{eq}}}\right)^{\!\!\!2}Y_N^2\!\right) \nn \\
-&\frac{\tilde{\Gamma}_{N\to {\rm SM}}}{H(m_\chi)} 
\,x \left(Y_N -Y_N^{\text{eq}}\right) + \frac{\tilde{\Gamma}_{\phi \to \chi N}}{H(m_\chi)} 
\,x \left(Y_\phi - \frac{Y_\phi^{\rm eq}}{Y_\chi^{\rm eq} Y_N^{\rm eq}} \, Y_\chi Y_N\right) \nn \\
-&  \frac{\tilde{\Gamma}_{N \to \chi \phi}}{H(m_\chi)} 
\,x \left(Y_N - \frac{Y_N^{\rm eq}}{Y_\chi^{\rm eq} Y_\phi^{\rm eq}}\,  Y_\chi Y_\phi\right).
\end{align}
Here $Y_i \equiv n_i/s$, describing the actual number of particle $i$ per comoving volume, where $n_i$ being the number density, and the variable $x \equiv m_\chi /T$ .
The entropy density $s$ and Hubble parameter $H$ at the DM mass is
$$ 
s(m_\chi)= \frac{2 \pi^2 }{45} g_s\, m_\chi^3, \quad H(m_\chi)= \frac{\pi}{\sqrt{90}} \frac{\sqrt{g_*}}{M_{pl}} m_\chi^2, $$  
and $Y_i^{\text{eq}}$ is the equilibrium number density of $i$-th particle given by
\begin{align}
\label{eq:Yi_eq}
Y_i^{\text{eq}}&\equiv\frac{n_i^{\text{eq}}}{s} =\frac{45}{2\pi^4} \sqrt{\frac{\pi}{8}}\left( \frac{g_i}{g_s}\right) \left({\frac{m_i}{T}}\right)^{3/2} e^{-\frac{m_i}{T}}\,. 
\end{align}
The parameters $g_{s, *}$, counting the number of relativistic degrees of freedom are temperature-dependent and we use the values at $T\sim m_\phi$ or $m_\chi$ when the final DM density is frozen in.  In most cases we have the standard value of $g_{s,*} = 106.75$. 
The internal degrees of freedom $g_{\chi,N}=2$ for the two Majorana particles $\chi$, $N$ and $g_\phi=1$ for $\phi$ being the real scalar.

Note that the Boltzmann equations in~\eqref{eq:dYDM}-\eqref{eq:dYN} are written for generality and all terms can not be present simultaneously due to the kinematic constraints. We provide the expressions of all the thermal averaged annihilation cross sections and the decay widths in Appendix~\ref{sec:appendix}. 

We are now ready to discuss in details the dynamics of the DM genesis attaining the observed relic density $\Omega_{\rm DM} h^2 =0.1199 \pm 0.0022$ \cite{planck18}  via freeze-in mechanism. 
We assume $\chi$ as the primary DM candidate, that is, $m_{\chi}< m_{\phi}$. However, we include also  the situation of $\phi$ as a decaying DM which is readily realizable in a certain scenario of our framework. The DM relic density is then calculated as $\Omega_{\rm DM}= \Omega_\chi$ when $\phi$ decays early enough, or $\Omega_{\rm DM}= \Omega_\phi$ if $\phi$ survives till today.
Let us first note that the production of $\chi$ can arise from various annihilation processes: $\phi \phi \to \chi \chi$, $NN \to \chi \chi$, $\nu h \to \chi \phi$, and also from decay processes: $\phi \to \chi N$, $\phi \to \chi \nu$ or $N\to \chi \phi$. Another key process is the production of $\phi$ and $N$ from the thermal bath through, e.g., the Higgs portal coupling, $h h \to \phi \phi$ (or $h\to \phi\phi$),
 or the inverse decay of $N$, $\nu h \to N$. Thus, the freeze-in abundance of DM will be determined by the interplay of the three couplings $y_\nu$, $\kappa$ and $\lambda$.
Concentrating on the major process for the DM production, here, the freeze-in mechanism can be categorized into three simple scenarios. Of course, a more general situation is achievable through any mixture of the major processes.

\subsection{\sn{1}}
\label{sec:scen1}

The first scenario we consider is the case where the initial abundance of the dark sector particle $\phi$ essentially determines the DM density.  For this we take sizable coupling $\kappa$  leading to abundant production of $\phi$ through the Higgs-portal process: $hh\to \phi \phi$. 
The DM particle $\chi$ is then produced from the decay of $\phi$: $\phi \to \chi N$ or $\chi \nu$ depending upon the mass spectrum, or the equilibration process $\phi\phi \leftrightarrow \chi\chi$ accompanied by the $\phi$ decay.  The decay $\phi \to \chi \nu$ proceeds due to the mass mixing of the RHN $N$ with the SM neutrinos via $y_\nu$, and thus involves the product of couplings $y_\nu \lambda$.  
We illustrate these cases by choosing  three sets of parameters in Table~\ref{table:S1}.
The values of $m_\phi$ and $\kappa$ are chosen such that  the DM $\chi$ satisfies the observed relic density of the Universe.

\begin{figure}[!h]
	\begin{center}
			\includegraphics[width=0.37\linewidth]{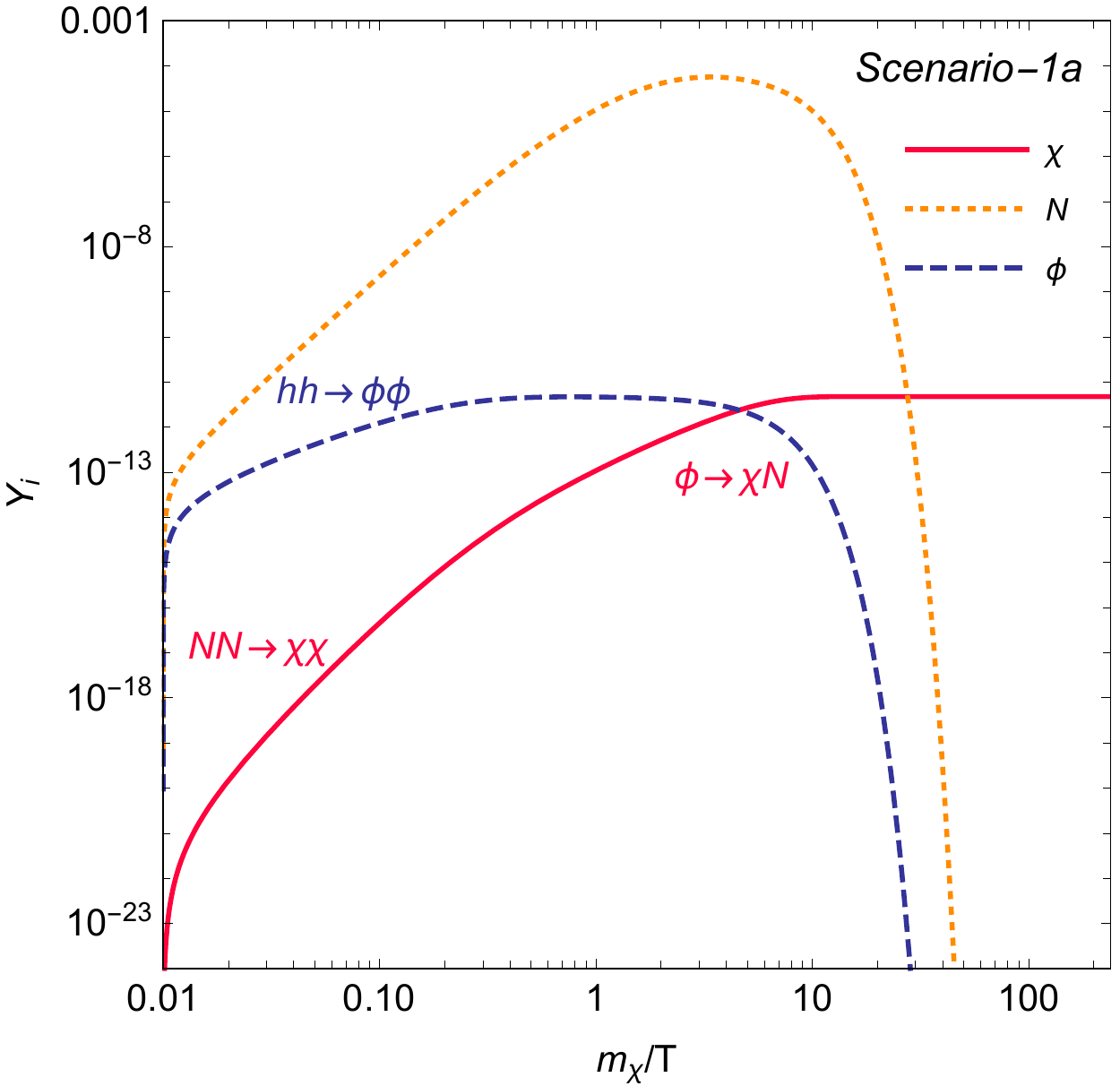} \hskip 20 pt \includegraphics[width=0.37\linewidth]{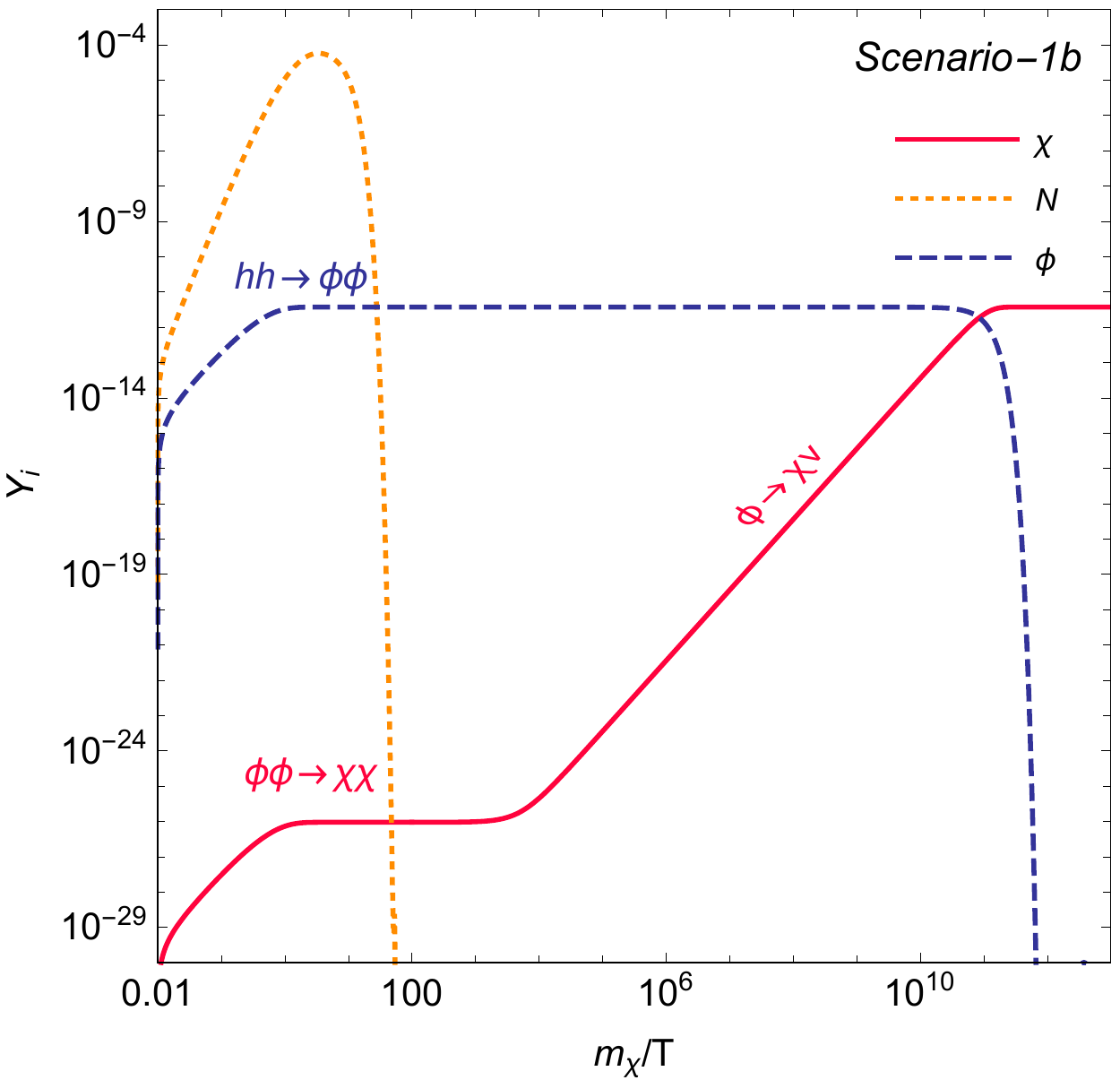}
			\includegraphics[width=0.37\linewidth]{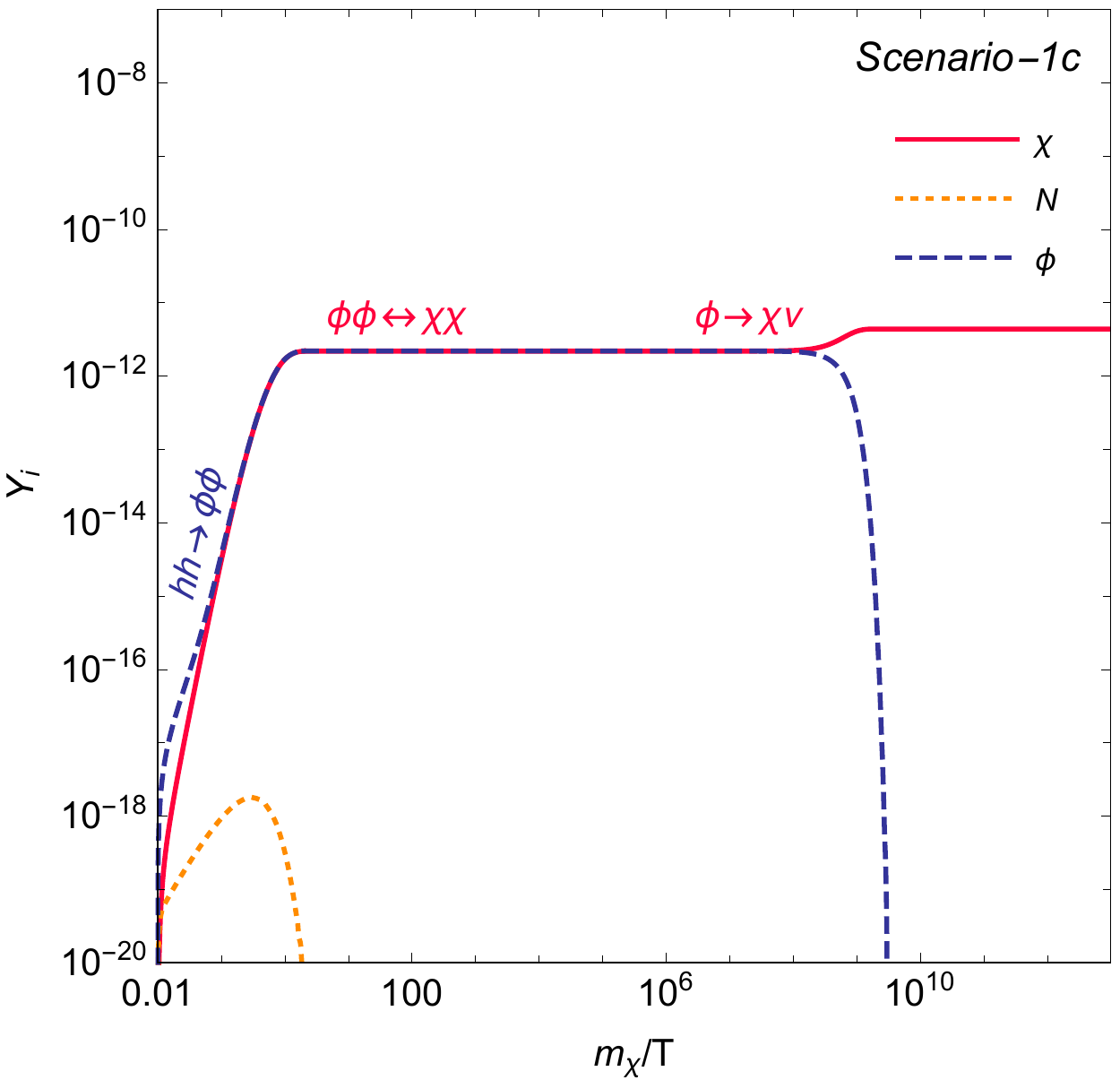} 
		\caption{The evolution of $\chi,~\phi$ and $N$ numbers per comoving volume for \sn{1a}, \sn{1b} and \sn{1c}. The observed DM relic density is satisfied for all three cases. }
		\label{fig:S1}
	\end{center}
\end{figure}

We show the evolution of the number densities of $\chi,\,\phi$ and $N$ in Fig.~\ref{fig:S1}.  The dominant processes are highlighted in the plots. In \sn{1a}, the initial production of $\chi$ mainly arise from the annihilation $N N \to \chi \chi$, which however becomes sub-dominant to the process $\phi\phi \to \chi\chi$ followed by $\phi \to \chi N $ ({\it 1a}), or $\chi \nu$ ({\it 1b}). 
For this, we require $\lambda \lesssim 10^{-6}$ not to maintain the process $\phi\phi \to \chi\chi$ in equilibrium and can allow it to take any smaller value. 

\begin{table}[h]
	\centering
	\renewcommand\arraystretch{1.1}
	\begin{tabular}{c| c  c  c | c  c c}
			\hline\hline
	  \multirow{2}{*}{ \textit{Scenario}  } 	& \multicolumn{3}{c}{Masses in \gev} & \multicolumn{3}{|c}{Couplings} \\
		            & $m_\chi$ & $m_N$ & $m_\phi$  & $y_\nu$ & $\kappa$ & $\lambda$ \\ \hline
	\textit{1a} & 100  & 200 & 500 & $10^{-8}$ & $4\times 10^{-11}$& $10^{-8} $ \\[1ex]
	\textit{1b} & 100  & 200 & 180 & $10^{-8}$ & $2\times 10^{-11}$& $10^{-10} $ \\[1ex]
	\textit{1c} & 100  & 500 & 250 & $2.5\times 10^{-12}$ & $10^{-12}  $& $10^{-4}$ \\
\hline\hline
\end{tabular}
\caption{The parameter choices for the three cases in \sn{1}.}
\label{table:S1}
\end{table}
Note that  the $\chi$ number density in \sn{1b} experiences a much slower rise compared to \sn{1a} as  the decay rate $\phi\to \chi \nu$ (involving $\lambda y_\nu $) is much smaller  than the $\phi \to \chi N$ rate (involving only $\lambda$). The decay $\phi\to \chi \nu$ in \sn{1b} has  a long lifetime $\tau_\phi \sim 10^{11}\,s$. We note that the coupling $\lambda y_\nu$ can be chosen much smaller to make the lifetime longer than the age of the Universe.  Such late decays 
will have important consequences in the neutrino experiments and will be addressed in Sec.~\ref{sec:DarkRad}.
 In \sn{1c}, we take $\lambda=10^{-4}$ which equilibrates $\phi\phi \leftrightarrow \chi\chi$ and later the decay of $\phi$ add the DM density. 
In \sn{1a}  and \sn{1b}, the initial number density of $N$ is higher compared to \sn{1c} due to larger $y_\nu$. The $\phi \to \chi N$ decay in \sn{1a} keeps producing $N$ until it finally becomes extinct.

\subsection{\sn{2}}
\label{sec:scen2}

Next one can conceive the case where the dark sector particles are predominantly produced via the annihilation of RHNs: $NN \to \chi \chi\, (\phi\phi)$. This requires sizable couplings $\lambda$ and $y_\nu$ as shown in Table~\ref{table:S2}.  
 We choose $\kappa=10^{-12}$  as in \sn{1c} so that the process $hh\to \phi\phi$ plays a minor role in the early stage of the evolution. However, we can keep $\kappa$ arbitrarily small as the Higgs portal process is essentially irrelevant in this scenario.

\begin{figure}[h]
	\begin{center}
		\includegraphics[width=0.39\linewidth]{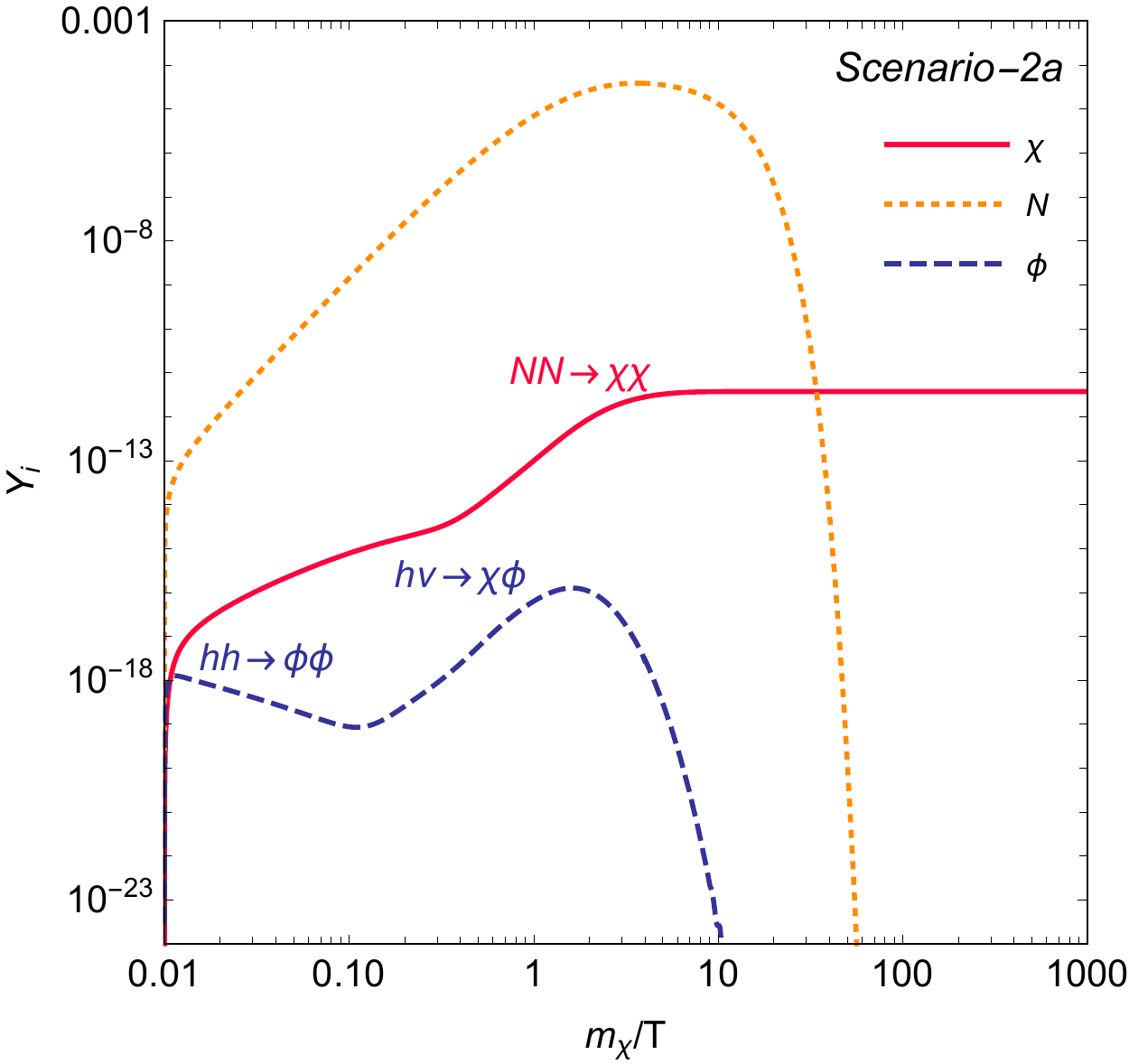} \hskip 20 pt \includegraphics[width=0.38\linewidth]{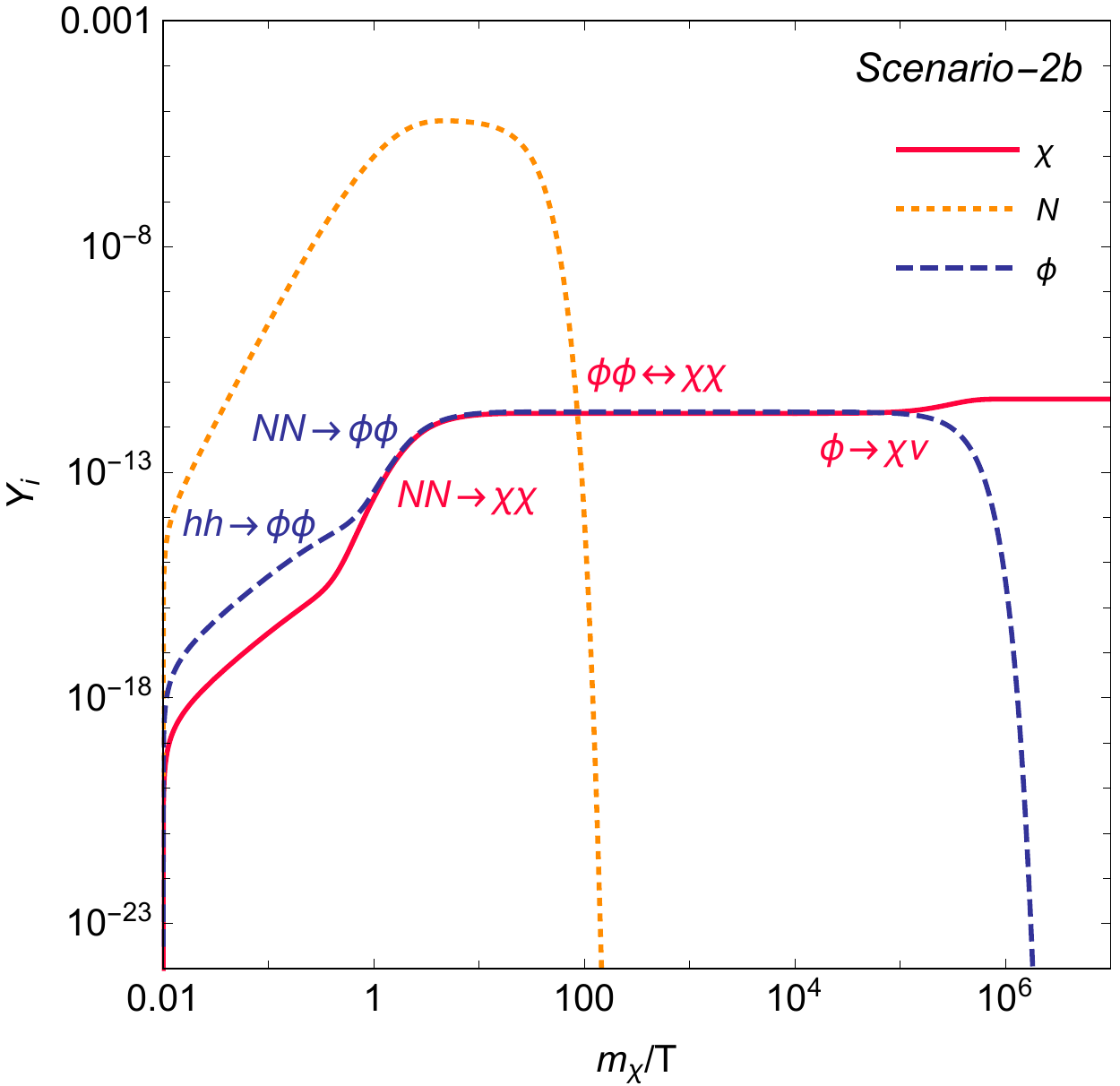} 
		\caption{The evolution of the  $\chi,~\phi$ and $N$ numbers per comoving volume for \sn{2a}  and \sn{2b}. The observed DM relic density is satisfied in both the cases.}
		\label{fig:S2}
	\end{center}
\end{figure}

In \sn{2a}, we take very heavy $\phi$ so that the process $NN \to \phi\phi$ is suppressed 
and  the fast decay of $\phi \to \chi N$ is allowed. As can be seen in the left panel of Fig.~\ref{fig:S2}, the abundance of $\phi$ is never comparable to that of $\chi$ although the process $h\nu\to \chi \phi$ can enhance the $\phi$ (and $\chi$) abundance a bit before it decays away.  The DM relic density is solely determined by the process $NN \to \chi \chi$.  On the other hand,
the two dark sector particles have similar masses in \sn{2b} and thus  their production rates  from the RHN annihilation are comparable and the late decay of $\phi\to \chi \nu$ is realized.  The evolution of the number densities, in this case, is presented in the right panel of Fig.~\ref{fig:S2}, where one can see a small excess of $\phi$  at the very beginning due to the  $hh\to \phi \phi$ process. However, the fast rise of the $N$ abundance dominates the production of $\phi$ and $\chi$ and their number densities become same due to the equilibration process $\phi\phi \leftrightarrow \chi\chi$ until the  $\phi \to \chi \nu$ decay occurs at the very late stage similar to \sn{1c}.

\begin{table}[h]
	\centering
	\renewcommand\arraystretch{1.1}
	\begin{tabular}{c| c  c  c | c  c c}
		\hline\hline
		\multirow{2}{*}{ \textit{Scenario}  } 	& \multicolumn{3}{c}{Masses in \gev} & \multicolumn{3}{|c}{Couplings} \\
		& $m_\chi$ & $m_N$ & $m_\phi$  & $y_\nu$ & $\kappa$ & $\lambda$ \\ \hline
		\textit{2a} & 100  & 200 & 500 & $8.0\times10^{-9}$ & $10^{-12}  $& $10^{-4} $ \\[1ex]
		\textit{2b} & 100  & 200 & 180 & $3.0\times 10^{-9}$ & $10^{-12} $& $10^{-4}$ \\		\hline\hline
	\end{tabular}
	\caption{The parameter choices for the two cases in \sn{2}.}
	\label{table:S2}
\end{table}

\subsection{\sn{3}}
\label{sec:scen3}

Finally,  the major production mechanism for the dark sector particles could be via the decay of the RHN.
This can be realized when the RHN is heavy enough to allow $N\to \chi \phi$, and the neutrino Yukawa coupling $y_\nu$ is relatively large to produce such a heavy  RHN in sufficient amount. In this case,  the coupling $\kappa$ becomes irrelevant for the DM production and can take any smaller values. However, $\lambda$ is crucial as increasing (decreasing) it reduces (enhances) the $\phi$ lifetime but it also reduces (enhances) the $N$ lifetime which on a contrary enhances (reduces) the $\phi$, and also $\chi$, number density via $N\to \chi \phi$. Hence,  for illustration, we take $\kappa =10^{-12}$ as before, and an optimal $\lambda$ to make $\phi$ very long-lived as presented in Table~\ref{table:S3}. We can even consider the case of  a decaying DM $\phi$ by making its lifetime longer than the age of the Universe. Such a long-lived $\phi$ leads to amusing observational consequences as will be discussed in the next section.  
\sn{3a} and \sn{3b} are very similar in nature except that  the latter deals with the mass spectrum in the PeV range. The relevant couplings  also exhibit the expected scaling in order to obtain the observed DM relic density.  We note that the lifetimes of $\phi$ for {\it 3a} and ${\it 3b}$ are $\tau_\phi \simeq 1\times 10^{10}\,s$ and $4.1\times 10^{12}\,s$, respectively, which are chosen to be around the matter-radiation equality time ($\sim 10^{12}\,s$) for the discussion in the 
succeeding section.

\begin{table}[h]
	\centering
	\renewcommand\arraystretch{1.1}
	\begin{tabular}{c| c  c  c | c  c c}
		\hline\hline
		\multirow{2}{*}{ \textit{Scenario}  } 	& \multicolumn{3}{c}{Masses in \gev} & \multicolumn{3}{|c}{Couplings} \\
		& $m_\chi$ & $m_N$ & $m_\phi$  & $y_\nu$ & $\kappa$ & $\lambda$ \\ \hline
			\textit{3a} & 100  & 341 & 241 & $10^{-7}$ & $10^{-12} $& $6.1\times 10^{-11}$ \\
			\textit{3b} &$1.0\times 10^6$ &$2.05\times 10^6$ &$1.05\times 10^6$ & $10^{-5}$&$10^{-12}$  &$2.4\times 10^{-11}$ \\
		\hline\hline
	\end{tabular}
	\caption{The parameter choices for the two cases in \sn{3}.}
	\label{table:S3}
\end{table}
\begin{figure}[!h]
	\begin{center}
		\includegraphics[width=0.38\linewidth]{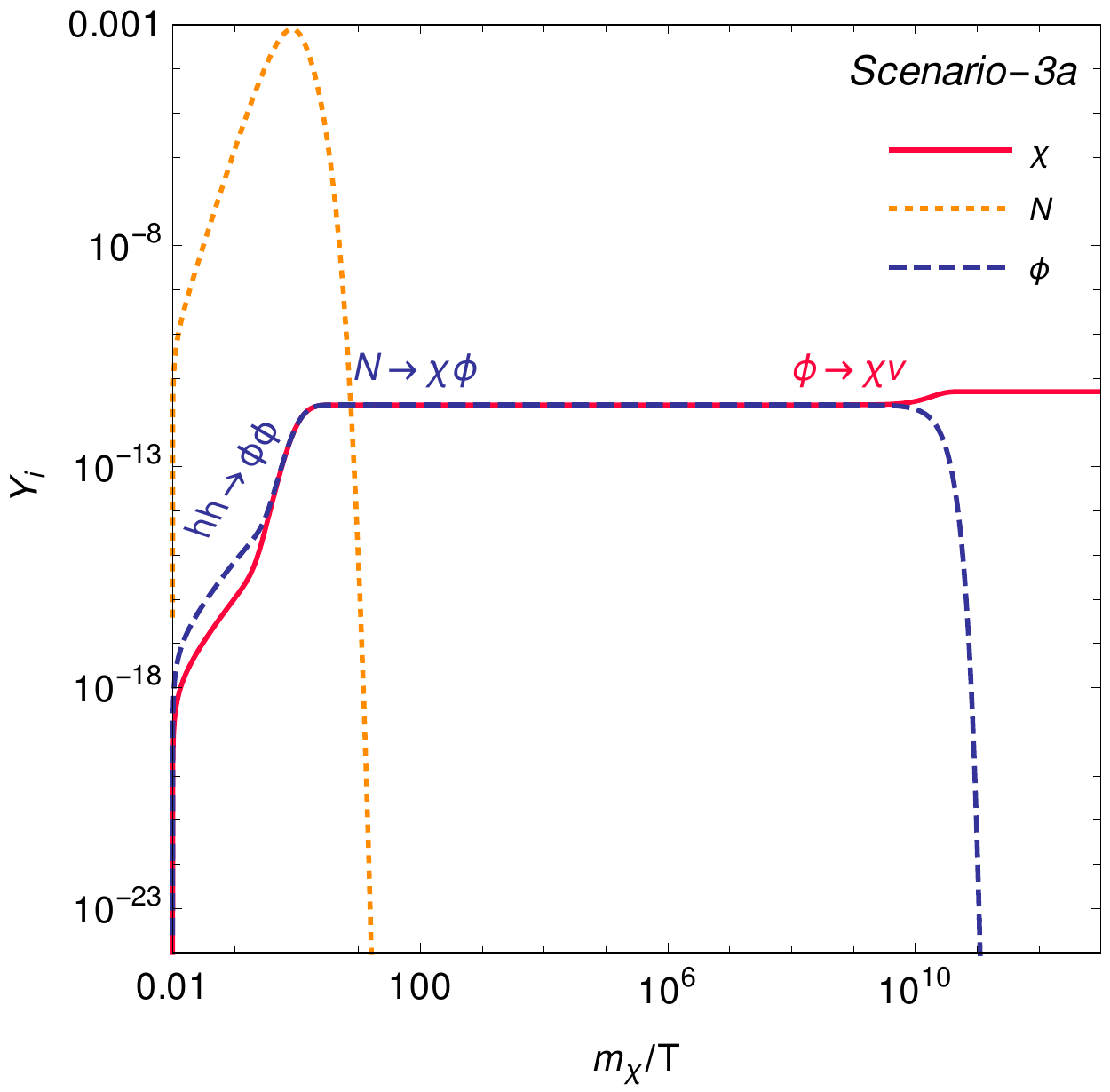} \hskip 20 pt
		\includegraphics[width=0.38\linewidth]{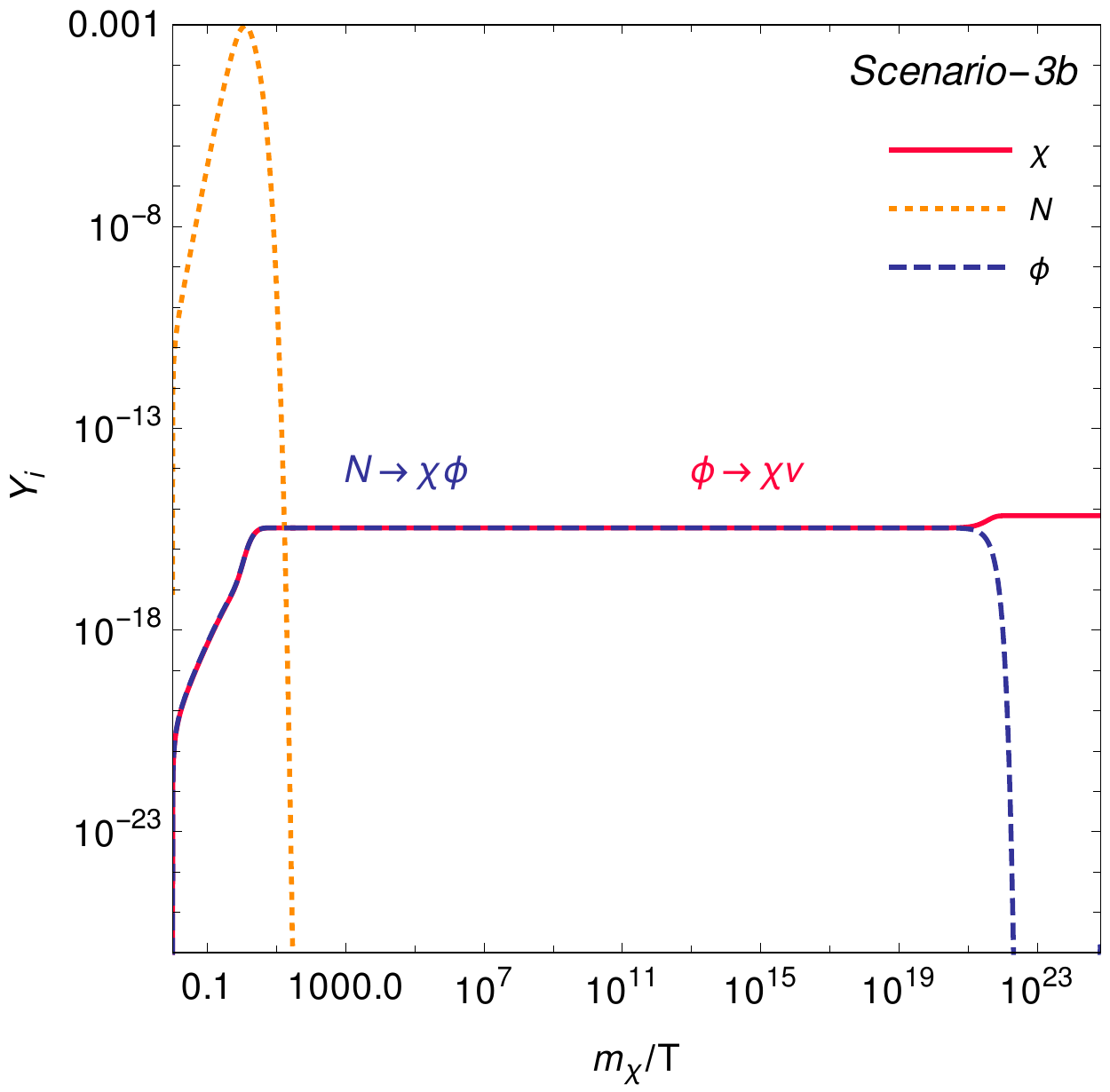} 
		\caption{The evolution of the $\chi,~\phi$ and $N$ number per comoving volume for \sn{3a} and \sn{3b}. The observed DM relic density is satisfied in both the cases.} 
		\label{fig:S3}
	\end{center}
\end{figure}
Figure~\ref{fig:S3} depicts the number density profiles. It can be seen that the $hh \to \phi \phi$ process dominates the $\phi$ production and then $N \to \chi \phi$ takes over enforcing the same density of $\chi$ and $\phi$ for a long enough time. The enhancement of the initial $N$ number density is attributed to large $y_\nu$ value. The final DM density is obtained after the very late decay $\phi \to \chi \nu$.

\section{Energetic neutrinos from dark sector}
\label{sec:DarkRad}

An interesting feature of the freeze-in scenarios of neutrino-portal DM is the source of  neutrinos arising from the decay of a dark sector particle, namely, $\phi \to \chi \nu$. This  is generated via the mixing of a RHN with a SM neutrino and its rate is proportional to the combination of feeble couplings $\lambda^2 y_\nu^2$ ~(see Eq.~\eqref{eq:Pdecaynu})  which is tiny as mentioned in the previous sections. It leads to a very late decay of $\phi$ to energetic neutrinos broadening the experimental scopes of the DM searches. Specific examples are provided in {\it Scenario-1b, 3a} and {\it 3b} where the parameters are chosen to have the lifetime of $\phi$, $\tau_\phi$, in the ballpark of the matter-radiation equality time. 
However, $\tau_\phi$ can be made arbitrarily large taking smaller $\lambda y_\nu$. 

Depending upon $\tau_\phi$, one can think of the following situations. 
More precise discussions will be made in the subsequent paragraphs.
\begin{itemize}
\item $ 1\, {\rm sec} \lesssim \tau_\phi \lesssim t_{\rm eq}$:  $\chi$ is the legitimate (stable) DM candidate, and the energetic neutrinos produced from the decay of $\phi$ are red-shifted away for $\tau_\phi \ll t_{\rm eq}$. Here $t_{\rm eq}= 5.11 \times 10^4\,$yrs is the matter-radiation equality time.

\item $t_{\rm eq} \lesssim \tau_\phi \ll t_0$: $\phi$ behaves like a  decaying DM and has disappeared by now.   The produced neutrinos as DR 
constrained by the CMB measurements, although red-shifted, are detectable at the neutrino experiments. 

\item $\tau_\phi \gtrsim t_0$:  $\phi$ is a decaying DM, and the energetic neutrino production puts stringent limit on $\tau_\phi \gg t_0$. Here, $t_0= 13.87$\,Gyr is the age of the Universe.
In this case, we have $\Omega_{\rm DM}=\Omega_\phi$ contrary to the previous cases with $\Omega_{\rm DM}=\Omega_\chi$. 
\end{itemize}

When $\tau_\phi \ll t_0$, the differential neutrino fluxes arriving at earth, $d{\varphi}_{\rm cos}/dE_\nu$,
originate  dominantly from the cosmological unclustered $\phi$ abundance and thus the red-shift information is crucial. The neutrinos emitted at the red-shift $1+z$ with the initial energy $E_0 $ from the decay $\phi \to \chi \nu$  is red-shifted to the observed energy $E_\nu $ at present where
\begin{equation}
 E_0 = \frac{m_\phi^2 - m_\chi^2}{2 m_\phi}~ {\rm and}~ E_\nu = \frac{E_0}{1+z}\,.
\end{equation}
  Then the observed neutrino flux can be obtained as
\begin{equation}
\label{eq:dfluxteq}
\frac{d\varphi_{\rm cos}}{dE_\nu}= \frac{n_\phi^0}{\tau_\phi} \int\limits_0^\infty dz\, \frac{e^{-t(z)/\tau_\phi}}{H(z)} \frac{dN}{dE_\nu}\,,
\end{equation}
where $n^0_\phi$ is the `present' $\phi$ number density if it were stable, and is the same as the produced neutrino and the DM $\chi$ number densities. 
Therefore, we have $n^0_\phi = \rho_{\rm DM}/m_\chi$ with $\rho_{\rm DM} = 0.126 \times 10^{-5}   \gev/{\rm cm}^3$ is the observed DM energy density. 
Recall that the Hubble parameter in the standard cosmology is given by 
\begin{equation}
H(z)  =   H_0 \sqrt{ \Omega_\Lambda + (1+z)^3 \Omega_{\rm m} + (1+z)^4 \Omega_{\rm r} }\, ,
\end{equation}
where $\Omega_\Lambda=0.6846$, $\Omega_{\rm m}=0.315$ and  $\Omega_{r}=9.265 \times 10^{-5}$ are  the dark energy, matter and radiation (CMB photons and neutrinos) fraction, respectively, and $H_0=100\,h$\,km/s/Mpc is the Hubble constant  with $h=0.6727$ \cite{planck18}.
Then, the cosmic time at red-shift $1+z$ is given by $
t(z) = \int_z^\infty dz' [(1+z') H(z')]^{-1} $ which can be well approximated as
\begin{align}
t(z) \approx \frac{4}{3 H_0} \frac{\Omega_{\rm r}^{3/2}}{\Omega_{\rm m}^2} \bigg[ 1- \left( 1- \frac{\Omega_{\rm m}}{2 (1+z) \Omega_{\rm r}} \right) \sqrt{1+  \frac{ \Omega_{\rm m}}{(1+z) \Omega_{\rm r}} }\bigg]\,,
\end{align}
in the most parameter space of our interest ($z>2$).
As the 2-body decay spectrum reads $dN/dE_\nu= \delta(E_\nu(1+z)-E_0)$, evaluating the red-shift integral Eq.~\eqref{eq:dfluxteq} we get the $E^2_\nu$-weighted flux
\begin{equation}
\label{eq:fluxCos}
\Phi_{\rm cos} \equiv  E_\nu^2 \frac{d\varphi_{\rm cos}}{dE_\nu}=E_\nu \frac{n_\phi^0}{\tau_\phi}  \frac{e^{-t(z)/\tau_\phi}}{H(z)} \theta(z)\,,
\end{equation}
where $z=E_0/E_\nu -1$.

The decay of dark sector particle to invisible relativistic components are constrained by the CMB measurements. While the bounds are stronger in case of the decay happening between today and the recombination time, the limits are somewhat relaxed when the dark sector particle decay before the recombination. 
From a recent update in the analysis of the impacts of the decaying DM on the CMB and the matter power spectrum~\cite{Poulin:2016nat}, we can draw interesting bounds on the parameter space of the portal under consideration. First of all,  a fraction of 10\% is tolerable at 95\% CL for the DM decaying before $\sim t_{\rm eq}$. This implies that the condition $\rho_{\rm DR}/\rho_{\rm DM} \lesssim 0.1$ at $t_{\rm eq}$
for the DR and DM components coming from the decay $\phi\to \nu \chi$. 
Considering the  red-shift property of DR and DM, we get the limit:
\begin{align}
\label{eq:CMB}
\frac{E_0}{m_\chi} \left( \frac{\tau_\phi}{t_{\rm eq}} \right)^{1/2} \lesssim 0.1\,,
\end{align}
for $\tau_\phi<t_{\rm eq}$.  
For the $\phi$ lifetime in the intermediate region; $t_{\rm eq} \lesssim \tau_\phi \ll t_0$,
the CMB can allow $E_0/m_\chi \lesssim 4\%$, or equivalently $m_\phi-m_\chi \lesssim 0.04 m_\chi$, in our case.
On the other hand, when the lifetime of decaying DM is greater than the age of the Universe, one finds the bound $\tau_\phi \gtrsim 170\,$Gyr at most~\cite{Poulin:2016nat}. We will see that much stronger bounds can be obtained from neutrino flux measurements in the relevant region of the parameter space.

\begin{figure}[ht]
	\begin{center}
		\includegraphics[width=0.8\linewidth]{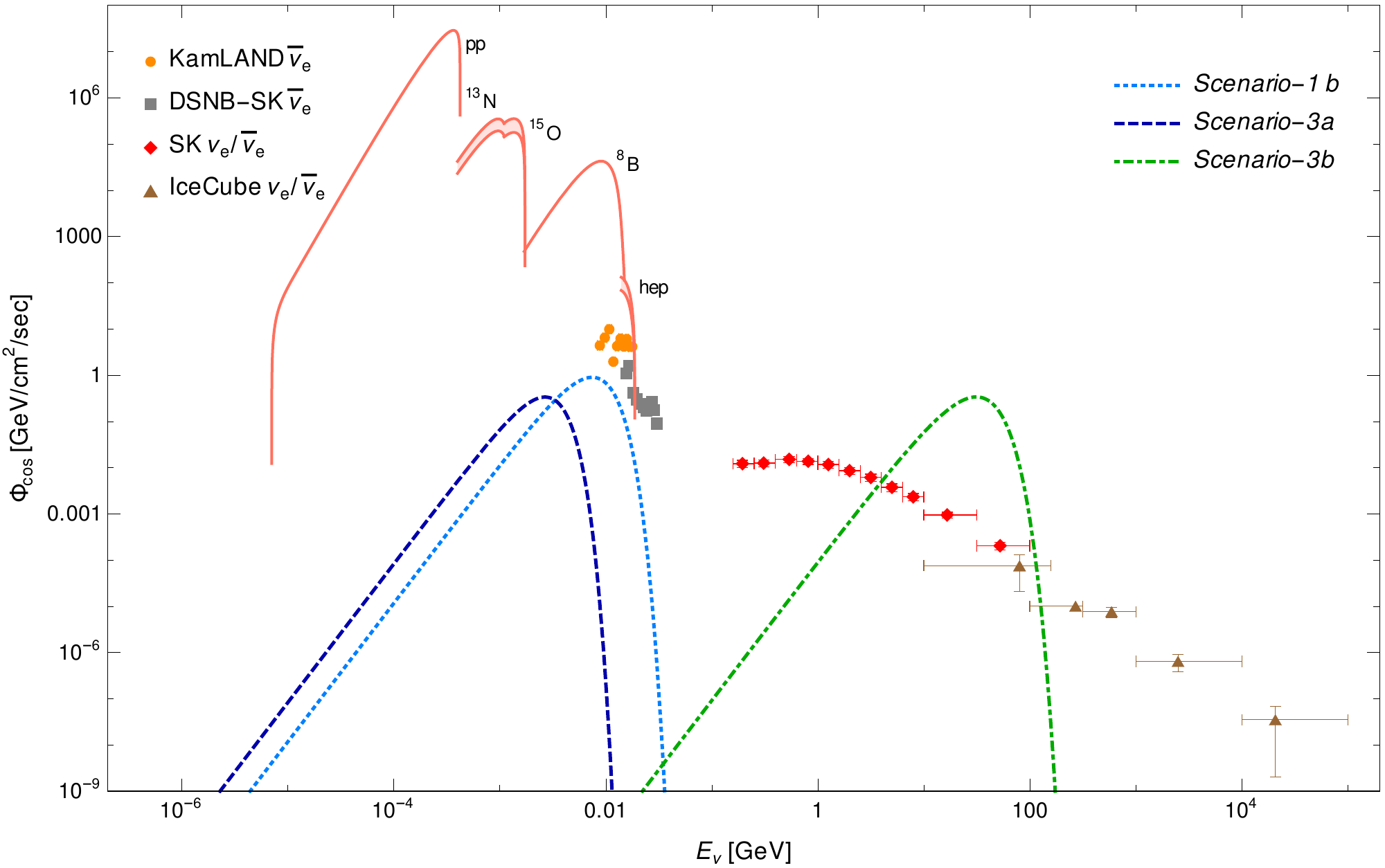} 
		\caption{The electron (anti)neutrino fluxes for \sn{1b}, \sn{3a} and \sn{3b} compared with the DSNB measurements at KamLAND~\cite{KamLAND}, SK~\cite{Bays:2011si} and atmospheric data from SK~\cite{superk} and IceCube~\cite{icecube}  together with the flux predictions of the standard solar model~\cite{Vitagliano:2019yzm}. \sn{1b} and \sn{3a} are allowed whereas \sn{3b} is ruled out by these measurements. }
		\label{fig:BPflux}
	\end{center}
\end{figure}

To illustrate the cosmic neutrino fluxes predicted in our model, let us consider \sn{1b}, \sn{3a} and \sn{3b} with $\tau_\phi$ around $t_{\rm eq}$ as discussed in Secs.~\ref{sec:scen1} and \ref{sec:scen3}. These scenarios satisfy the above mentioned bound from CMB measurements (Eq.~\eqref{eq:CMB}) and in Fig.~\ref{fig:BPflux} we show the diffuse supernova neutrino background (DSNB) flux of the electron anti-neutrinos from $\phi\to \chi \nu$ compared with the KamLAND~\cite{KamLAND} and SK~\cite{Bays:2011si} data. The measurements of solar neutrino flux is found to be consistent with the standard solar model predictions which are shown in red solid curves for the several different nuclear production processes in the Sun~\cite{Vitagliano:2019yzm}.  
It can be seen that \sn{1b} and \sn{3a} are allowed by the data whereas the highly energetic neutrinos arising in \sn{3b} are ruled out by the observed events at SK~\cite{superk} and IceCube~\cite{icecube} for atmospheric electron (anti)neutrino.
To be consistent with the observed flux data we need to take smaller $\tau_\phi$ and $E_0/m_{\chi}$ values which shift the peak of the spectrum to the left and downward, respectively.

When $\tau_\phi \gtrsim  t_0$,  $\phi$ serves as the (decaying) DM candidate and the local DM cluster becomes the major source of energetic neutrinos.  
The resulting galactic neutrino flux at earth is given by
\begin{equation}
\frac{d \varphi_{\mathrm{gal}}}{d E_\nu}=\frac{  e^{-t_{0} / \tau_\phi}}{\tau_\phi m_\phi} \frac{d N}{d E_\nu} \times R_{\mathrm{sol}}\, \rho_{\mathrm{sol}}\,\langle J\rangle\,,
\end{equation}
where $R_{\rm sol}= 8.33$\,kpc and $\rho_{\mathrm{sol}}=0.3\,\gev/{\rm cm}^3$ are the distance to the galactic center and the DM density at the position of the Earth and $\langle J\rangle\simeq 2.1$ is the average over the $J$-factor calculated from a NFW profile. To compare with the data, we also use the $E^2_\nu$-weighted flux:
\begin{equation}
\label{eq:fluxGal}
\Phi_{\rm gal} = E_\nu^2 \frac{d\varphi_{\rm gal}}{dE_\nu}=E_\nu^2 \frac{  e^{-t_{0} / \tau_\phi}}{\tau_\phi m_\phi} \frac{d N}{d E_\nu} \times R_{\mathrm{sol}}\, \rho_{\mathrm{sol}}\,\langle J\rangle\,.
\end{equation}
Here $d N/d E_\nu = \delta (E_\nu -E_0)$ coming from the $\phi\to \chi \nu$ decay and thus we have the line spectrum at the observed neutrino energy $E_0$.

\begin{figure}[ht]
	\begin{center}
		\includegraphics[width=0.44\linewidth]{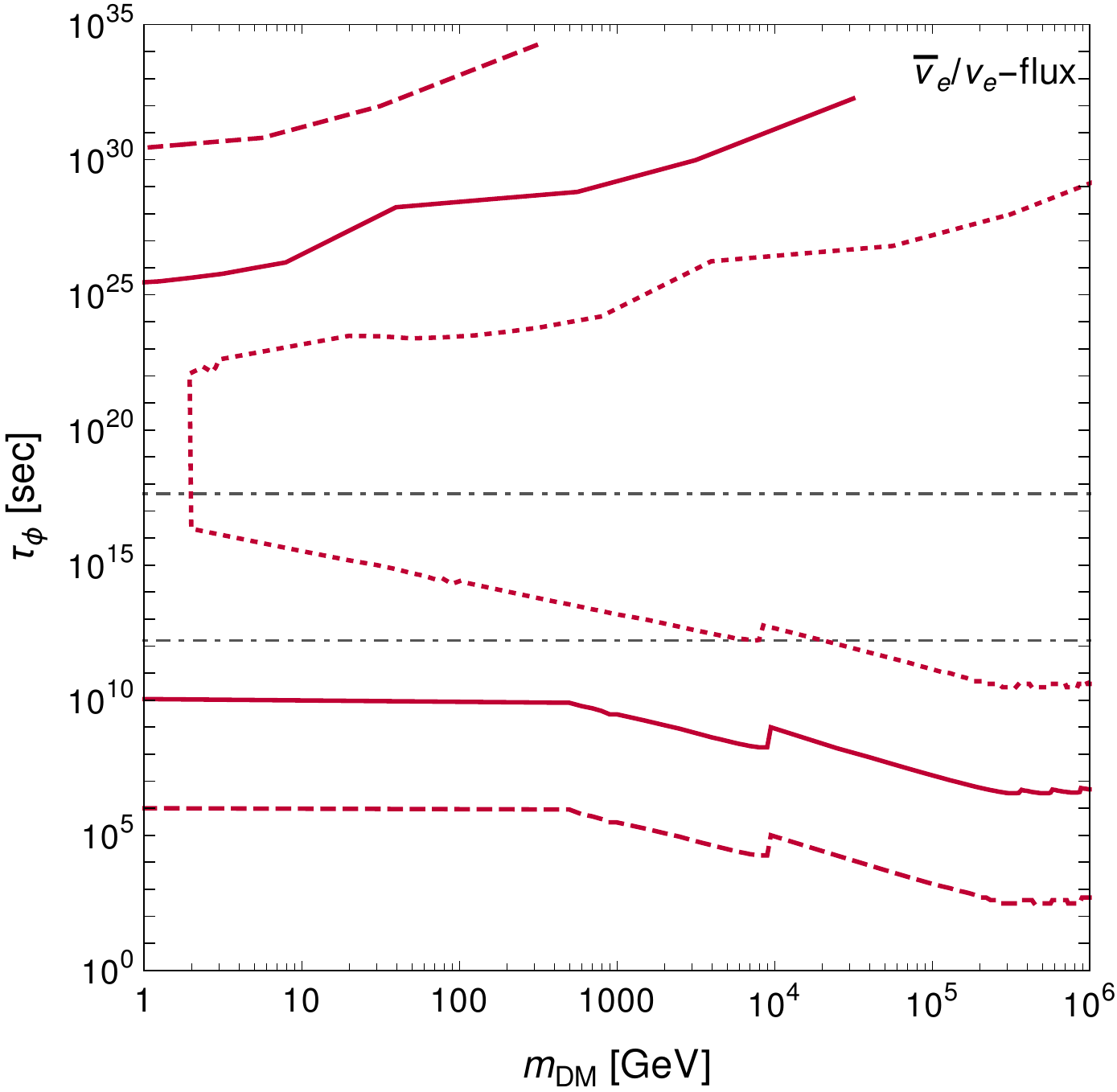} \hskip20pt
		\includegraphics[width=0.44\linewidth]{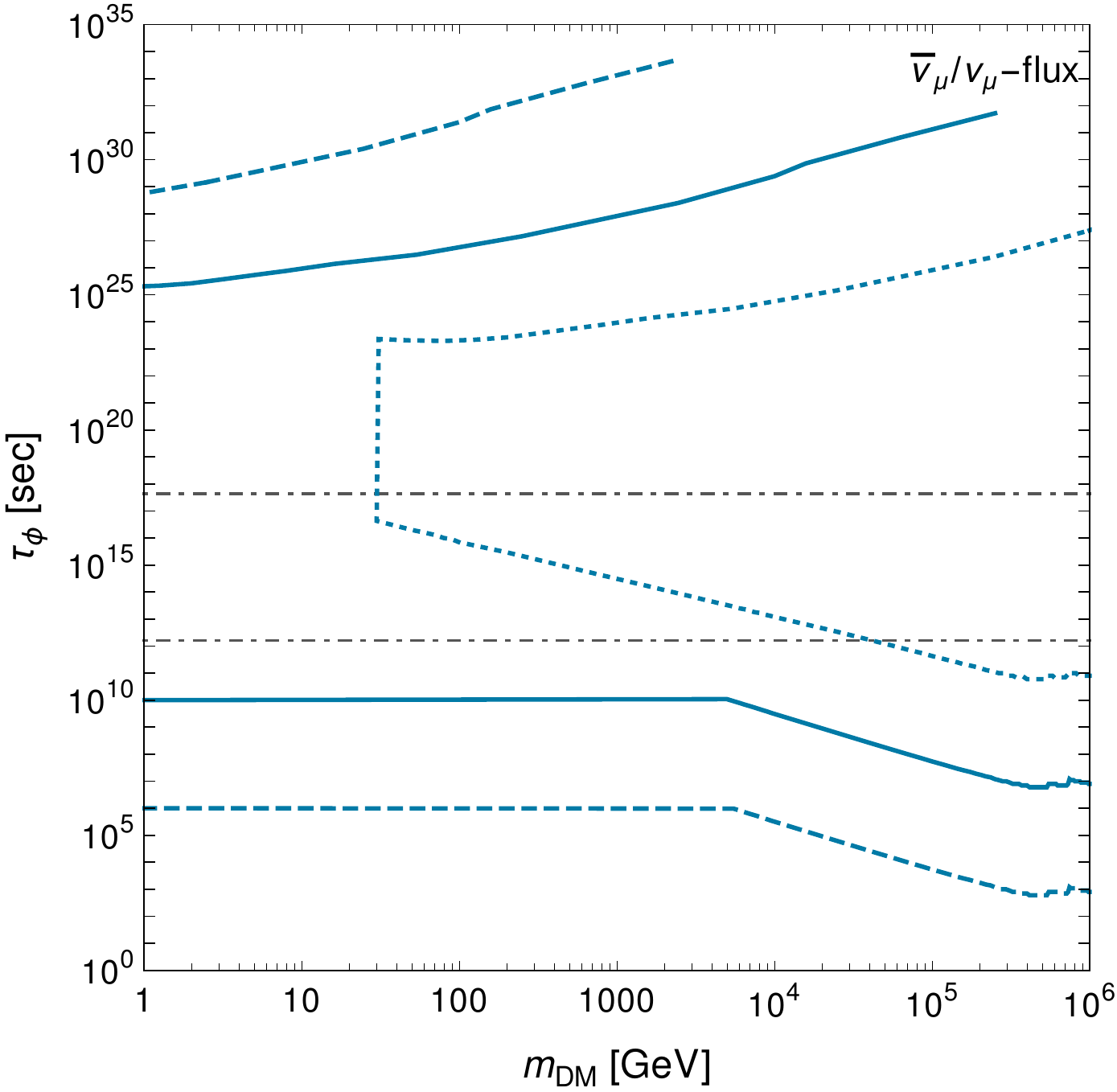} %
		\vskip10pt \includegraphics[width=0.45\linewidth]{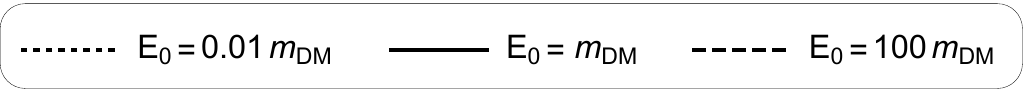} 
		\caption{The allowed region in $m_{\rm DM}-\tau_\phi$ plane from the KamLAND~\cite{KamLAND}, SK\cite{superk,Bays:2011si} and  IceCube \cite{icecube} measurements of DSNB and atmospheric (anti)neutrino flux, for three different choices of $E_0$. The left (right) panel corresponds to the $\bar{\nu}_e/\nu_e\, (\bar{\nu}_\mu/\nu_\mu)$ flux. The regions bounded by any two curves of a particular $E_0$ value are forbidden by the measurements.   The upper (lower) horizontal dot-dashed line refers to  the age of the Universe (matter-radiation equality time). }
		\label{fig:flux}
	\end{center}
\end{figure}

The two types of neutrino fluxes can be sensitive to the various detectors on earth dedicated to look for neutrinos and their existing data can constrain the portal parameters as we have already seen in the preceding paragraph. Thus we perform a generic study to explore the model parameter space where we have three input parameters $m_\chi,~m_\phi$ and $\tau_\phi$. We trade $m_\phi$ for the initial energy of the neutrino i.e., $E_0$ and compute the flux for a wide range of the DM mass $m_{\rm DM}$ and $\tau_\phi$ with different $E_0$ choices. The flux spectrum is then compared with the available data from KamLAND~\cite{KamLAND}, SK \cite{superk,Bays:2011si} and  IceCube \cite{icecube} experiments for the entire range of $E_\nu$ for both electron and muon flavors. 
The DSNB data from KamLAND~\cite{KamLAND} and SK~\cite{Bays:2011si} focus in the $\mathcal{O}$(10\,MeV) energy region for electron anti-neutrino.
The SK data for atmospheric neutrinos ranges from $\mathcal{O}(10^{-1}\gev)-\mathcal{O}(100\gev)$ whereas IceCube is sensitive to the higher energy regime i.e., $\mathcal{O}(100\gev)-\mathcal{O}(10\,{\rm TeV})$.
In the case of the galactic neutrino flux, the bounds are obtained by taking the line spectrum of Eq.~\eqref{eq:fluxGal} smeared with a 5\% width of $E_0$~\cite{Cui:2017ytb} at $E_\nu=E_0$.

The results  for three different choices of $E_0$ are depicted in Fig.~\ref{fig:flux} where the regions between the upper and lower curves for a given $E_0$ are forbidden. More tight constraints are obtained for the electron neutrino due to the DSNB data as well as slightly stronger observed bounds from the atmospheric neutrinos. One can see dislocations in the lower lines appearing at $m_{\rm DM} \approx 9$ TeV, which reflect the unavailability of measurement  between the DSNB and atmospheric neutrino data ($E_\nu \approx 30\,$MeV$- 0.1\gev$),  This also opens up small allowed islands slightly above the lower curves for the DM mass range $5-9\,$TeV which, however, are not shown in the figure. The flat nature of the curves in the lower half of the plots corresponds to the upper bound on $\tau_\phi$ from the CMB measurements (Eq.~\eqref{eq:CMB}). On the other hand, in the region of $\tau_\phi>t_0$ the obtained limits are stronger by two to three orders than the CMB bound demanding $\tau_\phi \gtrsim 170\,$Gyr at most.

\section{Conclusions}
\label{sec:concl}

A minimal extension of the SM in context of a feebly interacting neutrino portal DM has been explored. The portal dealing with two dark sector particles: a Majorana fermion $\chi$ and a real scalar $\phi$, interact with the SM sector via a RHN $N$. In addition, the real scalar also couples to the Higgs boson permitting the Higgs portal scenario. Due to the feeble nature of the couplings, these three beyond the SM particles are never in thermal equilibrium and their abundances are produced via various freeze-in processes. We comprehensively analysis the freeze-in productions categorizing into three main scenarios which are further split into sub categories originating from the kinematics. We assume $\chi$ as the primary DM candidate in our analysis and also discuss the consequences of the decaying DM $\phi$. We start with solving the coupled Boltzmann equations and describe the Higgs-portal scenario (in Sec.~\ref{sec:scen1}), where $\phi$  is produced via $hh\to \phi \phi$ and $\chi$ then attains the DM abundance through $\phi\to \chi N$ or $\phi\to \chi \nu$ processes. Next we highlight the neutrino portal case where the RHN is responsible for the DM production via $NN\to \chi\chi$ process in Sec.~\ref{sec:scen2} or through its decay $N\to \chi\phi$ in Sec.~\ref{sec:scen3}.

A very interesting feature of this portal is its detectability at the neutrino experiments dedicated to observe the neutrino sky over a large range of energies.
The energetic neutrinos arise in this portal from the very late decay $\phi \to \chi \nu$ generated through the mixing of the RHN with the SM neutrinos in the usual seesaw framework.
In particular, we find the DSNB and atmospheric neutrino flux measurements at 
KamLAND~\cite{KamLAND}, SK \cite{superk,Bays:2011si} and  IceCube \cite{icecube} experiments are sensitive  probe to the model parameter space. Two different situations emerge for the additional neutrino sources: the lifetime of $\phi$ typically around the matter-radiation equality time or greater than the age of the Universe (which dictates the decaying DM case). By calculating the cosmic and galactic flux distribution, and imposing the bounds from CMB and the matter power spectrum, in Sec.~\ref{sec:DarkRad}, we highlight the allowed region in the DM mass and $\phi$ lifetime plane. As an illustration, the benchmark scenarios satisfying the observed relic abundance are shown to be allowed (\sn{1b} and \sn{3a}) or forbidden (\sn{3b}), assuring the detectability of the portal with the future sensitivity of the experiments.

\section*{Acknowledgments}
E.J.C. acknowledges support  from InvisiblesPlus RISE No.~690575. P.B. wants to thank SERB project (CRG/2018/004971) for the support towards this work.
The work of R.M. has been supported by the Alexander von Humboldt Foundation through a postdoctoral research fellowship. 

\appendix

\section{Cross sections}
\label{sec:appendix}
In this section we provide the expressions for cross sections and decay widths involved in the coupled Boltzmann equations Eqs.~\eqref{eq:dYDM}--\eqref{eq:dYN}. First we quote the expressions of the thermal averaged cross section where we used $s-$wave approximation.
\begin{align}
\label{eq:sigvNNXX}
&\langle\sigma v \rangle_{NN \to \chi\chi} = \frac{\lambda^4 \left(m_\chi + m_N \right)^2}{16 \pi \left( m_N^2 +m_\phi^2 -m_\chi^2 \right)^2} \left(1- \frac{m_\chi^2}{m_N^2}\right)^{1/2} \hspace{-0.5cm}, \\
\label{eq:sigvPPXX}
&\langle\sigma v\rangle_{\phi\phi\to\chi\chi} =\frac{\lambda^4 \left(m_\chi + m_N \right)^2}{2 \pi \left( m_\phi^2-m_\chi^2 + m_N^2 \right)^2} \,  \left(1- \frac{m_\chi^2}{m_\phi^2}\right)^{3/2}\hspace{-0.5cm}, \\
\label{eq:sigvPPNN}
&\langle\sigma v\rangle_{\phi\phi\to N N} =\frac{\lambda^4 \left(m_\chi + m_N \right)^2}{2 \pi \left( m_\phi^2 + m_\chi^2 - m_N^2 \right)^2} \,  \left(1- \frac{m_N^2}{m_\phi^2}\right)^{3/2}\hspace{-0.5cm}, \\
\label{eq:sigvNNPP}
&\langle\sigma v\rangle_{NN \to \phi\phi } =\frac{\lambda^4 m_N^2}{8 \pi \left( m_N^2 + m_\chi^2 - m_\phi^2 \right)^2} \,  \left(1- \frac{m_\phi^2}{m_N^2}\right)^{3/2}\hspace{-0.5cm}, \\
\label{eq:PPhh}
&\langle\sigma v\rangle_{\phi\phi\to hh} = \left(\!1\!- \frac{m_h^2}{m_\phi^2}\right)^{\!\!\!1/2} \!\Bigg[ \frac{1}{64 \pi m_\phi^2} \left( 2\, \kappa + \frac{6 \,\kappa \,m_h^2 (4 m_\phi^2 - m_h^2)}{(4 m_\phi^2 - m_h^2)^2 + m_h^2 \Gamma_h^2}\right)^2\!\!+ \frac{\kappa^4 v^4 }{2 \pi m_\phi^2 \left( 2 m_\phi^2 -m_h^2 \right)^2} \Bigg], 
\end{align}
\begin{align}
\label{eq:PPSM}
&\langle\sigma v\rangle_{\phi\phi\to h\to {\rm SM}} = \frac{\sqrt{2}\kappa^2 v^2 G_F}{\pi\left((4 m_\phi^2 - m_h^2)^2 + m_h^2 \Gamma_h^2\right)} \times \Bigg[ 3 m_t^2\left(1- \frac{m_t^2}{m_\phi^2}\right)^{\!\!3/2}\!\!\!\! + 2 m_\phi^2\left(1- \frac{m_W^2}{m_\phi^2}\right)^{\!\!1/2}\!\!\!\!  \nn \\
&\hspace{2.2cm}+  m_\phi^2\left(1- \frac{m_Z^2}{m_\phi^2}\right)^{\!\!1/2}\Bigg]\,, \\
\label{eq:coXP}
&\langle\sigma v\rangle_{\chi\phi\to h\nu} = \frac{\lambda^2 y_\nu^2}{8 \pi } \frac{(m_N+m_\chi+m_\phi) (m_\chi(m_N+2 m_\chi)-m_\chi m_\phi+m_\phi^2 )}{(m_\chi+m_\phi)[((m_\chi+m_\phi)^2-m_N^2)^2 + m_N^2 \Gamma_N^2]} \left(1-\frac{m_h^2}{(m_\chi+m_\phi)^2} \right)^{\!1/2}\hspace{-0.5cm}.
\end{align}

The thermal decay width is denoted as
\begin{equation}
\tilde{\Gamma}_{X\to AB} =\frac{K_1(m_X/T)}{K_2(m_X/T)}\, \Gamma(X\to AB)\,,
\end{equation}
where the expressions for each individual process are 
\begin{align}
\label{eq:Pdecay}
\Gamma(\phi \to \chi  N) =& \frac{\lambda^2}{8 \pi}
\frac{m_\phi^2 - (m_\chi + m_N)^2}{m_\phi^3}
\bar\lambda^{1/2}\left(m_\phi^2,m_\chi^2, m_N^2 \right )\;, \\
\label{eq:Pdecaynu}
\Gamma(\phi \to \chi  \nu) =& \frac{1}{16 \pi} \frac{\lambda^2 y_\nu^2 v^2}{m_N^2}
m_\phi  \left(1- \frac{m_\chi^2}{m_\phi^2}\right)^2\hspace{-0.1cm}, \\
\label{eq:NdecayDM}
\Gamma(N \to \chi \phi) =& \frac{\lambda^2}{16 \pi}
\frac{(m_\chi + m_N)^2-m_\phi^2 }{m_N^3}
\bar\lambda^{1/2}\left(m_\phi^2,m_\chi^2, m_N^2 \right ) \;, \\
\label{eq:Ndecayh}
\Gamma(N\to h \nu)=&\Gamma(N\to h \bar{\nu}) 
= \frac{y_\nu^2 m_N}{64\pi} \left(1- \frac{m_h^2}{m_N^2}\right)^2, \\ 
\Gamma(N\to \ell^- W^+)= &\Gamma(N\to \ell^+ W^-)
= \frac{y_\nu^2 m_N}{32\pi} \left(1- \!\frac{m_W^2}{m_N^2}\right)^2\!\! \left(1+ 2 \frac{m_W^2}{m_N^2}\right), \\ 
\label{eq:NdecayZ}
\Gamma(N\to Z \nu ) = &\Gamma(N\to Z \bar{\nu})
=  \frac{y_\nu^2 m_N}{64\pi}  \left(1- \frac{m_Z^2}{m_N^2}\right)^2 \!\! \left(1+ 2 \frac{m_Z^2}{m_N^2}\right)\,,
\end{align}
with $\bar \lambda (x,y,z) = x^2 +y^2 +z^2 -2xy -2xz-2yz$ is the K\"allen function.

\end{document}